\documentclass[conference]{IEEEtran}
\IEEEoverridecommandlockouts
\usepackage[hidelinks]{hyperref}
\usepackage{cite}
\usepackage{physics}
\usepackage{amsmath,amssymb,amsfonts}
\usepackage{algorithmic}
\usepackage{graphicx}
\usepackage{textcomp}
\usepackage{xcolor}
\usepackage{ulem}
\definecolor{bs}{rgb}{1.0, 0.44, 0.37}

\def\BibTeX{{\rm B\kern-.05em{\sc i\kern-.025em b}\kern-.08em
    T\kern-.1667em\lower.7ex\hbox{E}\kern-.125emX}}
\begin{document}
\title{A True Random Number Generator for Probabilistic Computing using Stochastic Magnetic Actuated Random Transducer Devices}

\author{\IEEEauthorblockN{Ankit Shukla$^{1, a}$, Laura Heller$^{1, b}$, Md Golam Morshed$^{2, c}$, Laura Rehm$^{3, d}$, Avik W. Ghosh$^{2, e}$, \\
Andrew D. Kent$^{3, f}$, Shaloo Rakheja$^{1, g}$}
\IEEEauthorblockA{$^{1}$\textit{Holonyak Micro and Nanotechnology Laboratory}, University of Illinois at Urbana-Champaign, Urbana, IL USA \\
$^{2}$\textit{Department of Electrical and Computer Engineering},
University of Virginia, Charlottesville, VA USA \\
$^{3}$\textit{Center for Quantum Phenomena, Department of Physics},
New York University, New York, NY USA \\
\{$^a$ankits4, $^b$lheller4, $^g$rakheja\}@illinois.edu, \{$^c$mm8by, $^e$ag7rq\}@virginia.edu, \{$^d$laura.rehm, $^f$andy.kent\}@nyu.edu}}

\maketitle

\thispagestyle{plain}
\pagestyle{plain}

\begin{abstract}
Magnetic tunnel junctions (MTJs), which are the fundamental building blocks of spintronic devices, have been used to build true random number generators (TRNGs) with different trade-offs between throughput, power, and area requirements. MTJs with high-barrier magnets (HBMs) have been used to generate random bitstreams with $\lesssim$ 200~Mb/s throughput and pJ/bit energy consumption. A high temperature sensitivity, however, adversely affects their performance as a TRNG. Superparamagnetic MTJs employing low-barrier magnets (LBMs) have also been used for TRNG operation. Although LBM-based MTJs can operate at low energy, they suffer from slow dynamics, sensitivity to process variations, and low fabrication yield. In this paper, we model a TRNG based on medium-barrier magnets (MBMs) with perpendicular magnetic anisotropy. The proposed MBM-based TRNG is driven with short voltage pulses to induce ballistic, yet stochastic, magnetization switching. We show that the proposed TRNG can operate at frequencies of about 500~MHz while consuming less than 100~fJ/bit of energy. In the short-pulse ballistic limit, the switching probability of our device shows robustness to variations in temperature and material parameters relative to LBMs and HBMs. Our results suggest that MBM-based MTJs are suitable candidates for building fast and energy-efficient TRNG hardware units for probabilistic computing.
\end{abstract}

\begin{IEEEkeywords}
True random number generation, Magnetic tunnel junctions, Energy-efficient computing, Probabilistic switch, Process variability, Spintronics
\end{IEEEkeywords}

\section{Introduction}
Random number generators (RNGs) are used in a wide variety of applications, such as cryptography, hardware security, Monte-Carlo simulations, and more recently in stochastic computing and brain-inspired computing\cite{stip2014, alaghi2013, rangarajan2021, misra2022}.
Commonly, mathematical or computational algorithms, referred to as pseudo-random number generators (PRNGs), are used to generate random sequences. These sequences are, however, completely deterministic and can be regenerated if the initial state (referred to as ``seed'') is known. PRNGs are, therefore, not suitable for cryptography or hardware security applications. Also, the generated sequences have long-range correlations, which could lead to errors in Monte-Carlo simulations and thus compromise the integrity of the solution~\cite{stip2014, misra2022}.

True random number generators (TRNGs), on the other hand, are hardware elements that utilize non-deterministic phenomena, such as thermal noise, oscillatory jitters, radioactive decay, among others, to generate secure stochastic bitstreams~\cite{stip2014, fu2021}.
Most existing TRNGs, however, either require complex and power consuming post-processing circuits to extract the random bits, or are too prone to temperature and device variations. Some of them also suffer from poor throughput, and as a result fail to meet the demands of emerging data-driven applications~\cite{stip2014, fu2021}.
Therefore, CMOS-compatible TRNGs that are fast, energy-efficient, compact, tunable, scalable for on-chip integration, and also robust to temperature and fabrication process variations are required. 

\begin{figure*}[t!]
\centering
\includegraphics[width = 2\columnwidth]{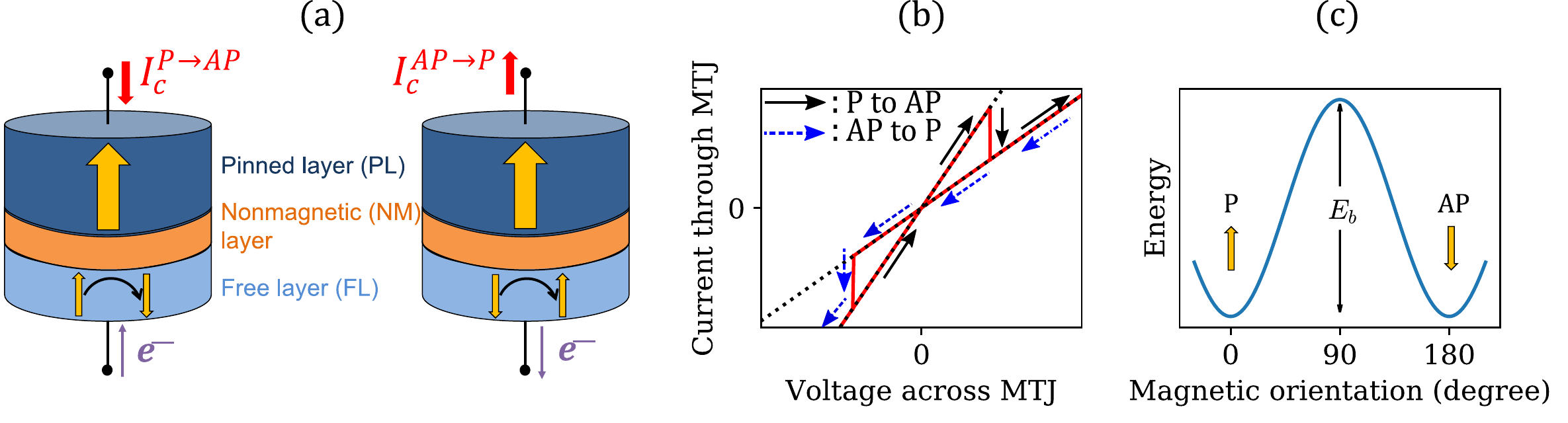}
\caption{(a) Spin-transfer-torque-driven switching in an MTJ. Current $I_c^\mathrm{P\to AP}$ ($I_c^\mathrm{AP\to P}$) switches the magnetization of the FL antiparallel (parallel) to that of the PL. 
(b) Current through the MTJ as a function of the applied voltage. The solid red lines show the behavior of an MTJ in the case of switching from parallel (P) to antiparallel (AP), and vice-versa. On the other hand, the dotted black lines (overlaid on the red lines) show the behavior of the MTJ if no switching had occurred.
(c) Macrospin model showing the energy barrier $E_b$ separating the P and AP states.}
\label{fig1}
\end{figure*}

Magnetic tunnel junctions (MTJs) are two-terminal spintronic devices which consist of a thin non-magnetic (NM) layer sandwiched between two ferromagnetic layers---a ``pinned-layer'' (PL) whose magnetization is fixed, and a ``free-layer'' (FL) whose magnetization can be reoriented by an input spin current or magnetic field~\cite{Kent2015}.
In spin current-driven MTJs, electrons flowing into the FL can switch the magnetization direction if sufficient angular momentum is transferred to this layer.
Electrons flowing from the FL (PL) into the PL (FL) switch the magnetization of the FL antiparallel (parallel) to that of the PL, as shown in Fig.~\ref{fig1}(a).
The relative magnetic orientations of the PL and FL can then be transduced into an electrical signal, as shown in Fig.~\ref{fig1}(b).
Parallel (antiparallel) magnetic orientations of the PL and FL corresponds to a low (high) resistance, and therefore a high (low) current is detected across the MTJ.
These binary resistance states are separated by an energy barrier $E_b$, as shown in Fig.~\ref{fig1}(c), and can be used to represent one bit of information, viz., bit `0' for low resistance and bit `1' for high resistance state, respectively.
The energy barrier in a ferromagnet (FM) depends on its material properties and the volume, and determines its thermal stability.
If $E_b$ is low and comparable to the thermal energy $k_BT$, where $k_B$ is the Boltzmann constant and $T$ is the temperature, the magnetization of FL fluctuates freely between the two states~\cite{vodenicarevic2017}.
Such FMs are referred to as low-barrier magnets (LBMs).
On the other hand, if $E_b > 40~k_BT$, the two magnetization states are stable and could be used for long-term (more than 10 years) storage~\cite{Kent2015}. These FMs are referred to as high-barrier magnets (HMBs).

Traditionally, MTJs have been used to realize non-volatile (NV) magnetic memory devices, as well as radio-frequency (RF) oscillators.
Moreover, they are compatible with CMOS back-end-of-line fabrication processes~\cite{Kent2015}, which makes them an attractive technology for enabling future energy-efficient computing platforms~\cite{guo2021}.   
The MTJ operation is stochastic in nature as the magnetization dynamics of a FM is strongly influenced by thermal effects.
On the one hand, these thermal effects lead to challenges such as write-errors in NV memories~\cite{Kent2015} and line-width broadening in RF oscillators~\cite{chen2016}.
On the other hand, stochastic MTJ operation can be used for developing electrically tunable TRNGs~\cite{fu2021}. For example, by controlling the pulse width and the amplitude of the applied input voltage or current, the switching probability of the FL can be modified~\cite{fukushima2014}. 

In this paper, we study a new type of fast, energy-efficient, and robust spintronic TRNG, relying on the stochasticity present in medium-barrier magnets (MBMs) with perpendicular magnetic anisotropy (PMA) that are used as the FL of an MTJ~\cite{rehm2023}. 
Typically, the energy barrier of MBMs is in the range of 20-40~$k_BT$.
The stochastic response of this TRNG is tuned by applying voltage pulses of appropriate amplitude and duration. 
Our simulations performed using 45-$\mathrm{nm}$ CMOS PDK~\cite{stine2007} show that the TRNG can operate at a frequency of 500~$\mathrm{MHz}$ with energy consumption of 92 $\mathrm{fJ/bit}$, while also being relatively insensitive to both process and temperature (PVT) variations. The TRNG presented in this paper offers high robustness, more than 2.0$\times$ higher frequency while also consuming 3.0$\times$ lower power compared to previous spintronic TRNG~\cite{rangarajan2017}.
The key contributions of this work are as follows.
\begin{itemize}
    \item[(i)] Description of the physics and operation and Monte-Carlo simulations of the MBM TRNG.
\item[(ii)]  Quantification of the process, voltage, and temperature robustness of the MBM TRNG and performance benchmarking against prior works.
    \item[(iii)] MBM TRNG circuit design in 45-$\mathrm{nm}$ CMOS node for timing-accurate analyses.
\end{itemize}

\section{Prior Work}\label{prior}
Previously, 1/f noise in transistors, shot noise associated with tunneling current in heavily-doped Zener diodes, accumulated jitter in ring oscillators, and random telegraph noise (RTN) in MOSFETs, have been exploited to build CMOS-compatible TRNGs with different pros and cons. 
Optics-based sources, such as lasers, entangled photons, and superconducting circuits have also been explored as prospective TRNG candidates.
They can be used to generate random sequences at 60~$\mathrm{Gbps}$ but require complex post-processing and low temperatures~\cite{li2012, hai2004, sugiura2010}.

Recently, stochastic switching behavior in oxide-based resistive elements, ferroelectric materials, magnetic materials, and phase-change-based materials, has been leveraged as source of true random bitstreams. 
Among these emerging devices, magnetic devices have become the front-runner for TRNG technology owing to their high throughput and significantly lower energy cost compared to other materials~\cite{fu2021}. 
In addition, MTJ-based TRNGs can be built using either HBMs~\cite{fukushima2014, rangarajan2017} or LBMs~\cite{vodenicarevic2017, parks2018}.
Finally, the FL used in both the HBM- or LBM-based MTJs could have either an in-plane anisotropy~\cite{rangarajan2017, vodenicarevic2017} or a perpendicular anisotropy~\cite{fukushima2014, parks2018}. 
These wide range of choices have enabled the development of CMOS-compatible and scalable MTJ-based TRNGs~\cite{fu2021}.

\subsection{Non-magnetic TRNGs}
While CMOS-based TRNGs are quite mature, they typically rely on 1/f noise in large number of transistors (more than 100) as entropy sources. 
However, the output bitstream is correlated, and the scalability of this TRNG is limited.
Large number of transistors also lead to area and power overheads. For example, a bistable CMOS-based TRNG fabricated in 45-nm CMOS technology operated at $2.4~\mathrm{GHz}$ but consumed $3~\mathrm{pJ/bit}$ energy and occupied $4000~\mathrm{\mu m^2}$ circuit area~\cite{mathew2012}. 
Another CMOS ring oscillator-based TRNG that utilized accumulated jitter to make the design more robust to noise injection consumed $23~\mathrm{pJ/bit}$ energy and occupied $375~\mathrm{\mu m^2}$ area while generating bits at the rate of $23~\mathrm{MHz}$~\cite{yang2014}. 
Zener diode-based TRNGs, which rely on shot noise associated with tunneling currents, suffer from large area and power overhead since they require post-processing to amplify their weak output signal, and reduce correlation~\cite{stip2014}. 
CMOS-compatible TRNGs utilizing RTN in MOSFETs and metal-insulator-metal memristors have limited bitrate ($\leq$ 3.33~$\mathrm{Hz}$ for a 65-$\mathrm{nm}$ RTN TRNG\cite{mohanty2017}), are generally not stable over time due to complex electrostatic interactions between trapped charges and ions, and are highly sensitive to electrode/insulator materials. Also, the measured RTN signal is quite weak. 

Non-magnetic oxide-based resistive elements, ferroelectrics, and phase-change-based TRNGs typically require energy-intensive programming operations, while their throughput is limited to 40~$\mathrm{Mb/s}$~\cite{jerry2017}. 
Besides, the entropy of phase-change TRNGs is highly sensitive to the stoichiometry
crystal quality and temperature~\cite{piccinni2017}. 
The stochastic relaxation dynamics in metal-insulator transition oxides (e.g., VO$_2$) has been used to generate random signals but the signals are extremely
sensitive to the device temperature and the materials are not generally CMOS compatible~\cite{valle2022}.

\subsection{MTJs with high-barrier magnets}
Previous works on spintronic TRNGs have largely focused on HMBs.
In~\cite{fukushima2014, choi2014}, antidamping-like torque was used to realize a TRNG with HBMs, while in~\cite{kim2015, rangarajan2017}, precessional magnetization dynamics are excited to generate random data stream.  
In these TRNGs, input spin current disturbs the FL from its stable state to a metastable state. An equal switching probability from the metastable state to one of the two stable equilibria points generates random bits~\cite{fu2021}.

TRNGs based on the antidamping switching mechanism are tunable but extremely sensitive to the amplitude and the duration of the input as well as to temperature variations~\cite{fukushima2014, rehm2023}. To address these challenges different circuit level modifications have been suggested. For example, in~\cite{choi2014, fu2021}, a 10-bit counter feedback circuit, which counts the number of `1's' in a random sequence and adjusts the input current pulse duration in real time to achieve 50\% switching probability, was presented.
This was followed by other feedback circuit implementations which replaced previously used high-gain analog amplifiers by their digital counterparts for better CMOS integration~\cite{fu2021, oosawa2015, vatajelu2016}. Here, either the current amplitude or its duration were adjusted to tune the switching probability close to 50\%.
These feedback circuits, however, introduce unwanted correlations in the bitstream. 
Recent, simulation results have showed that parallely connected MTJs could generate 
unbiased bitstreams while also avoiding complex feedback circuitry~\cite{qu2017, qu2018}.

TRNGs based on the precessional switching mechanism are faster and require lower energy, but have low tunability.
They are robust to small variations in input current amplitude and pulse duration, if the amplitude is above a certain threshold~\cite{kim2015, rangarajan2017, chen2018, fu2021}.
However, process and voltage variations in such TRNGs could lower the current amplitude below the threshold. 
It was shown through simulations that a feedback circuit with two counters could help in this regard~\cite{liu2018, fu2021}, but at the cost of unwanted correlations.

In all the above designs, post-processing using XOR gates would reduce bias in the bitstreams, however, it would incur additional area and power overhead, and lower the throughput~\cite{fukushima2014, rangarajan2017, chen2018}.

\subsection{MTJs with low-barrier magnets}
Recently, LBMs in the superparamagnetic limit have been used as energy-efficient sources of random numbers.
In superparamagnets, thermal noise passively flips the magnetization state of the FL from one stable state to the another without any input power. 
These fluctuations can then be sampled (by a small external read current) as a random bitstream~\cite{vodenicarevic2017, parks2018}.

An LBM-based TRNG with an in-plane magnetized FL was shown to consume only 2~$\mathrm{fJ/bit}$ of energy during the read process~\cite{vodenicarevic2017}. 
However, the generated bitstreams were biased and required 8 XOR operations to improve randomness.
This post-processing increased the energy cost to 20~$\mathrm{fJ/bit}$ and required a circuit area of $2~\mathrm{\mu m^2}$, while also lowering the bitrate (1.66~$\mathrm{kHz}$)~\cite{vodenicarevic2017}. 
Another TRNG with a perpendicularly magnetized FL consumed about 600~$\mathrm{fJ/bit}$ energy in order to generate random bits at 45~$\mathrm{kHz}$~\cite{parks2018}. 
Though further post-processing to improve the quality of bitstream increased power overhead, it was still 3$\times$ lower compared to CMOS designs, while the total circuit area of the TRNG was orders of magnitude smaller~\cite{parks2018}.  

In LBM-based TRNGs, the bitstream generation rate could be increased by either increasing the temperature or reducing the energy barrier between the two states~\cite{vodenicarevic2017}.
However, robust LBMs with barrier heights of just a few $k_B T$ are very difficult to reliably fabricate because in-plane magnetized FLs require near-perfect circular cross-sections, while perpendicularly magnetized FLs require precise cancellation of magnetic anisotropy and demagnetization fields. 
LBMs are also highly sensitive to process, temperature, and magnetic field variations, which reduces their robustness at the circuits and systems level~\cite{rehm2023}. 

\section{Stochastic Magnetic Actuated Random Transducer (SMART) TRNGs}
In this work, we present a TRNG based on Stochastic Magnetic Actuated Random Transducer (SMART) devices which utilize MBMs with perpendicular anisotropy as the FL within an MTJ structure~\cite{rehm2023}.
SMART TRNGs are fast, have an active write scheme, and are significantly less sensitive to PVT variations~\cite{rehm2023}, unlike LBM-based TRNGs that are slow, sensitive to variations and difficult to realize in practice~\cite{vodenicarevic2017, parks2018}. 
The actuation signal to the SMART device is provided by applying voltage pulses of specific amplitude and duration~\cite{rehm2023}. 
We demonstrate that by controlling the characteristics of the input signal, the switching probability of our TRNG can be tuned to 50\%. 
Its stochastic response is also found to be more robust to temperature fluctuations. 

\subsection{Numerical Model}\label{model}
To obtain the switching statistics and infer the relationship between switching probability and the characteristics of the input voltage pulses applied to the SMART device, the MBM with PMA is treated within the monodomain limit, wherein it has two stable states, as shown in Fig.~\ref{fig1}(c).
The probability of switching between these two states can be tuned to 50\% via the spin-transfer-torque (STT)~\cite{slonczewski1996, berger1996} resulting from the input voltage pulses. 
Therefore, the magnetization dynamics of the monodomain is numerically solved via the Landau-Lifshitz-Gilbert (LLG) equation, which includes the effect of thermal noise~\cite{ament2016, sun2016}. It is given as
\begin{equation}\label{sLLGS}
    \begin{split}
        \pdv{\vb{m}}{t} &= - \frac{\gamma \mu_0}{1+\alpha^2} \bigg [\qty(\vb{m} \cp \vb{H}_\mathrm{eff}) + \alpha \vb{m} \cp \qty(\vb{m} \cp \vb{H}_\mathrm{eff}) \bigg.\\  
        & \left. + \frac{\hbar}{2e}\eta\frac{G_\mathrm{P} V_\mathrm{ap}}{\mu_0 M_s \mathcal{V}}\qty(\vb{m} \cp \qty(\vb{m} \cp \vb{n}_{p}) - \alpha \qty(\vb{m} \cp \vb{n}_{p})) \right],
    \end{split}
\end{equation}
where $\vb{m}$ is the magnetization of the MBM normalized to the saturation magnetization $M_s$, $\alpha$ is the Gilbert damping constant, $V_\mathrm{ap}$ is the input voltage across the MTJ, $\eta$ is the spin polarization efficiency factor, $\vb{n}_{p}$ is the unit vector along the direction of the spin polarization, and $G_\mathrm{P}$ is the conductance of the P state.
Here, $\mathcal{V}$ is the volume of the FL layer, $\hbar$ is the reduced Planck’s constant, $\mu_0$ is the permeability of free space, $e$ is the elementary charge, and $\gamma = 17.6 \times 10^{10}$~T$^{-1}$~s$^{-1}$ is the gyromagnetic ratio.

The effective magnetic field, $\vb{H}_\mathrm{eff}$, includes contributions from the effective anisotropy field, $H_k$, acting along the easy axis, and a thermal field, $\vb{H}_\mathrm{th}$. Therefore, 
\begin{equation}\label{H_eff}
        \vb{H}_\mathrm{eff} =  H_k m_z \vb{z} + \vb{H}_\mathrm{th} (t), 
\end{equation}
where $H_k = \qty(\frac{2K_u}{\mu_0 M_s} - M_s)$, and the easy axis is assumed to be along the $\vb{z}$ direction.
The effective anisotropy field, $H_k$, depends on the field due to uniaxial magnetocrystalline anisotropy, $\frac{2K_{u}}{\mu_0 M_s}$, and the field due to shape anisotropy, which depends on $M_s$.
The thermal field is assumed to be a Langevin field that is spatially isotropic and uncorrelated in space and time~\cite{pinna2013}. 
The monodomain is assumed to be in equilibrium with a thermal bath at temperature $T$. 
The energy barrier of the MBM in the SMART device, can be normalized to the thermal energy as, $\Delta_{0} = \frac{\mu_0 M_s H_k \mathcal{V}}{2 k_B T}$.

\begin{figure}[t!]
\centering
\includegraphics[width = \columnwidth]{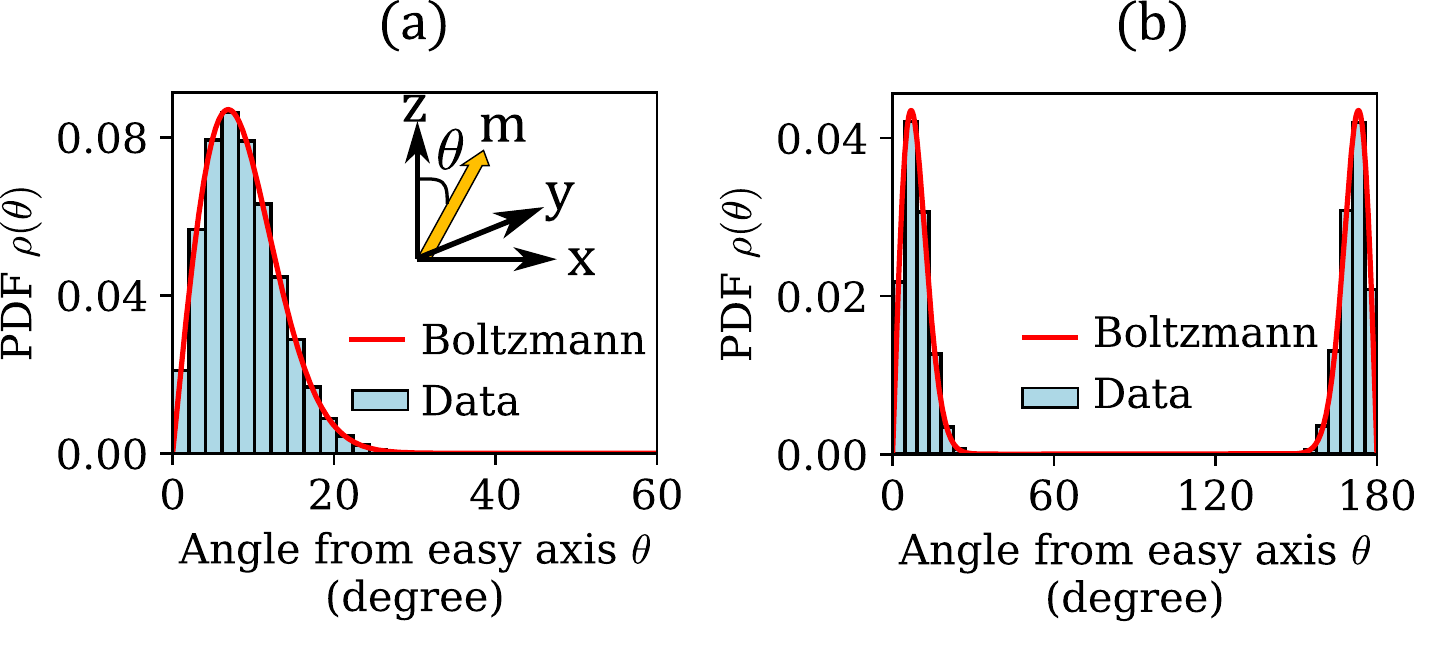}
\caption{(a) Equilibrium distribution of magnetization in the P state. 
The overlaid red curve corresponds to (\ref{distribution}) with $Z = F(\sqrt{\Delta_0})$.
(b) Steady-state bimodal distribution corresponding to a switching probability of 0.499. 
The overlaid red curve corresponds to (\ref{distribution}) with $Z = 2 F(\sqrt{\Delta_0})$.
In both the cases, the histograms consists of $10^5$ data points and were obtained from the solution of (\ref{sLLGS}).}
\label{fig2}
\end{figure}

\subsection{Effect of the input voltage on switching probability}
Equation~(\ref{sLLGS}) is numerically solved for an ensemble of $10^5$ spins using the Heun integration scheme with a 10~$\mathrm{fs}$ integration time step. 
Simulations are implemented in CUDA and run in parallel on GPUs for faster solutions.
The temperature of the system is assumed to be $T = 300$~$\mathrm{K}$, unless otherwise stated. Material parameters used for simulations are listed in Table~\ref{tab:param}.
In the absence of any input current and external field, the Boltzmann distribution of the magnetization angle, $\rho(\theta)$, of Fig.~\ref{fig2}(a) is obtained.
Numerical results are represented by the histogram whereas the red overlaid curve is given as~\cite{pinna2013}
\begin{equation}\label{distribution}
    \rho(\theta) = \frac{\sqrt{\Delta_0}}{Z}\exp^{-\Delta_0 \sin^2 \theta} |\sin \theta|,
\end{equation}
where $\theta = \cos^{-1}(m_z)$, and $\int_0^{\pi/2} \rho(\theta) d\theta = 1$. This leads to $Z = F(\sqrt{\Delta_0})$, the Dawson's integral. 

\begin{figure}[b!]
\centering
\includegraphics[width = 0.7\columnwidth]{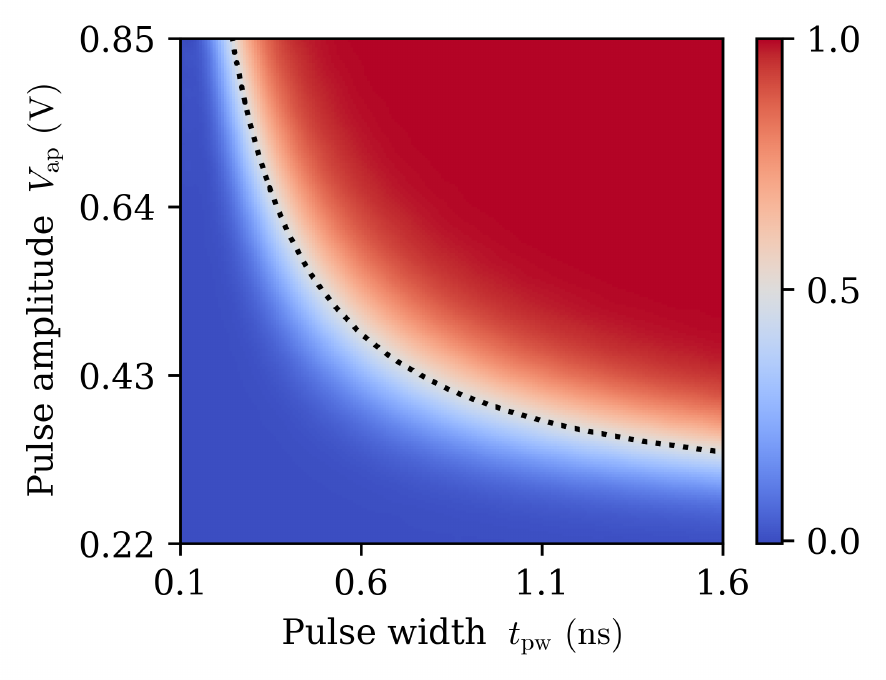}
\caption{Switching probability for pulse amplitude, $V_\mathrm{ap}$, ranging from 0.214~$\mathrm{V}$ to 0.854~$\mathrm{V}$ and pulse widths, $t_\mathrm{pw}$, ranging from 0.1~$\mathrm{ns}$ to 1.6~$\mathrm{ns}$. The dotted
black line corresponds to $P_\mathrm{sw} = 0.5$}
\label{pr_heatmap}
\end{figure}
The threshold voltage to switch the MBM deterministically, in the absence of any external field, is given as $V^\mathrm{th} = \frac{1}{\eta G_\mathrm{P}}\alpha \frac{2 e}{\hbar} \mu_0 M_s H_k \mathcal{V}$~\cite{sun2016}, which is equal to 0.374~$\mathrm{V}$ for the material parameters chosen.
Here, we explore the dependence of the magnetization switching probability, $P_\mathrm{sw}$, on the amplitude and the pulse width of the applied voltage. 
As shown in Fig.~\ref{pr_heatmap}, the amplitude of the applied voltage, $V_\mathrm{ap}$, is varied from 0.214~$\mathrm{V}$ to 0.854~$\mathrm{V}$, while its pulse width, $t_\mathrm{pw}$, is varied from 0.1~$\mathrm{ns}$ to 1.6~$\mathrm{ns}$.
The overlaid dotted black line in Fig.~\ref{pr_heatmap} represents the combination of $V_\mathrm{ap}$ and $t_\mathrm{pw}$ required to achieve $P_\mathrm{sw} = 0.5$, the ideal value for TRNG operation. 
It can be observed that lower pulse widths require higher pulse amplitudes, and vice-versa.
For voltage pulses leading to $P_\mathrm{sw} = 0.5$, the Boltzmann distribution of Fig.~\ref{fig2}(a) is driven out of equilibrium into the bimodal distribution of Fig.~\ref{fig2}(b).
Here, the overlaid red curve is given by~(\ref{distribution}) with $\int_0^{\pi} \rho(\theta) d\theta = 1$, which leads to $Z = 2F(\sqrt{\Delta_0})$.

\begin{figure}[t!]
\centering
\includegraphics[width = \columnwidth]{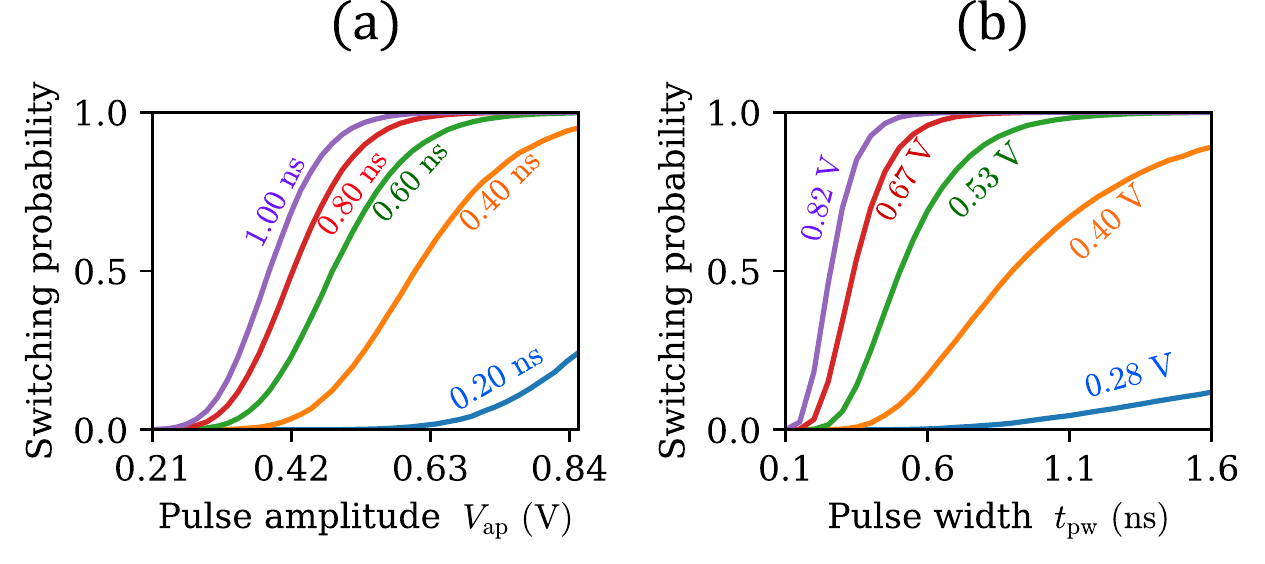}
\caption{ Switching probability as a function of the input voltage (a) pulse amplitude $V_\mathrm{ap}$, and (b) pulse width $t_\mathrm{pw}$. Probability of switching increases with both the amplitude and width of the pulse.}
\label{pr}
\end{figure}
Figure~\ref{pr} shows that $P_\mathrm{sw}$ is a monotonically increasing function of $V_\mathrm{ap}$ and $t_\mathrm{pw}$. 
Furthermore, both Figs.~\ref{pr_heatmap} and~\ref{pr}(a) indicate that the switching probability at $P_\mathrm{sw} = 0.5$ is more sensitive to the pulse amplitude for larger pulse widths.
On the other hand, it is more sensitive to the pulse widths for larger pulse amplitudes, as observed from Figs.~\ref{pr_heatmap} and~\ref{pr}(b).
Hence, a voltage pulse with an optimal $V_\mathrm{ap}$ and $t_\mathrm{pw}$ that minimizes this sensitivity is required for a robust TRNG circuit.

\begin{figure}[b!]
\centering
\includegraphics[width = \columnwidth]{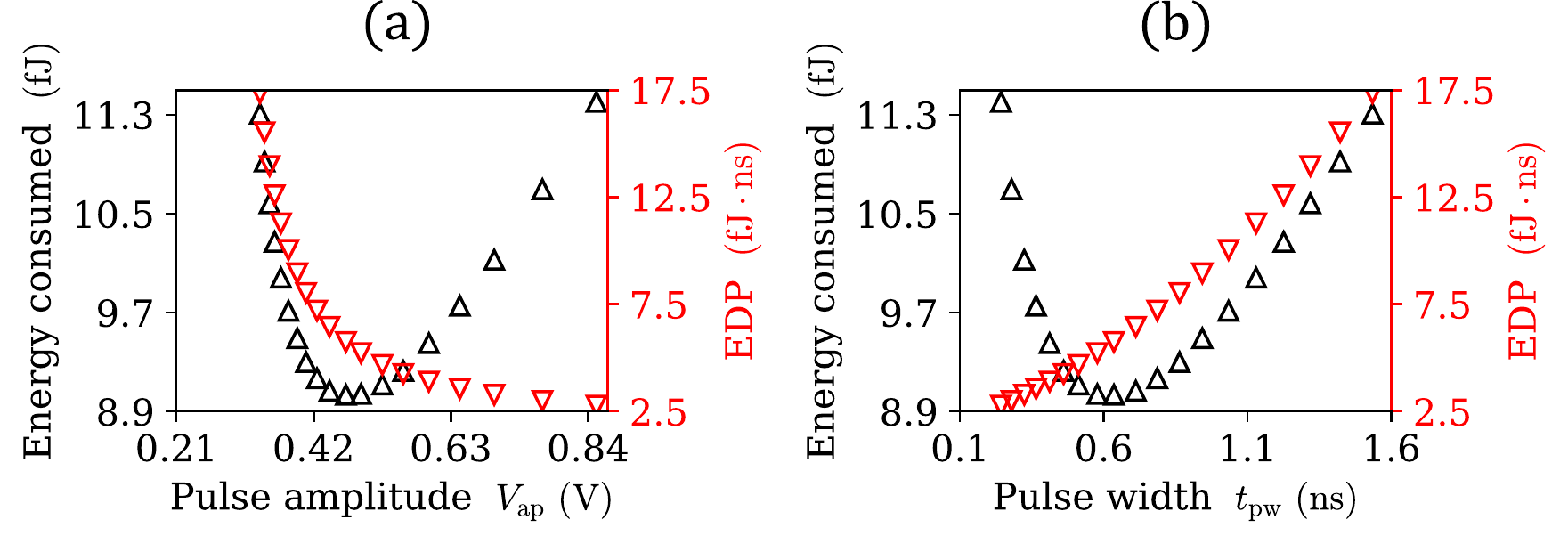}
\caption{Switching energy on the left axis and energy-delay product (EDP) on the right axis for $P_\mathrm{sw} = 0.5$ as a function of (a) Pulse amplitude 
(b) Pulse width.
Both the switching energy and EDP are minimized in the ballistic limit for short pulses.} 
\label{energy_delay}
\end{figure}

In order to build a fast and energy-efficient TRNG, both the energy consumed and the delay, or equivalently the energy-delay product (EDP) required to create the bimodal distribution should be minimal.
Figure~\ref{energy_delay} shows the energy consumed and the corresponding EDP required to switch the FL from P to AP state with $P_\mathrm{sw} = 0.5$.
The minimum energy is approximately 9~$\mathrm{fJ}$, and corresponds to $V_\mathrm{ap} = 0.48$~$\mathrm{V}$ and $t_\mathrm{pw} = 0.6$~$\mathrm{ns}$, in the ballistic regime~\cite{pinna2013}. The EDP at this operating point, therefore, is $5.45 \times 10^{-24}~\mathrm{J.s}$.
Generally, the EDP decreases as the pulse amplitude increases (Fig~\ref{energy_delay}(a)), while it increases as the pulse duration increases (Fig~\ref{energy_delay}(b)).
For EDP calculations, the delay is assumed to be the same as $t_\mathrm{pw}$.

\begin{table}[t!]
\caption{List of material parameters of the free-layer FM of the MTJ.}
\begin{center}
\begin{tabular}{|l|l|l|}
\hline
\textbf{Parameters}&\textbf{Description} & \textbf{Values}\\
\hline
$d$ \ ($\mathrm{nm}$) & Diameter & 15 \\
\hline
$t_f$ \ ($\mathrm{nm}$) & Thickness & 1.5 \\
\hline
$\mathcal{V}$ \ $\mathrm{(nm^3)}$ & Volume & $\pi d^2 t_f/4$ \\
\hline
$\alpha$ & Damping & $0.01$  \cite{kani2016}\\
\hline
$M_s \ \mathrm{(A/m)}$ & Saturation magnetization & $3 \times 10^5$ \cite{kani2016}\\
\hline
$K_u \ \mathrm{(J/m^3)}$ & Uniaxial anisotropy & $6 \times 10^5$ \cite{kani2016}\\
\hline
$\mathrm{RA} \ \mathrm{(\Omega \vdot \mu m^2)}$ & Resistance area product&  $2.74$ \cite{rehm2023}\\
\hline
$\mathrm{TMR} (V_\mathrm{ap} = 0)$ & Tunnel magnetoresistance& $45.4 \%$ \cite{rehm2023}\\
\hline
$R_\mathrm{P} \ \mathrm{(k \Omega)}$ & Resistance in P state&  $\mathrm{\frac{RA}{\pi d^2/4}} = 15.5$\\
\hline
$R_\mathrm{AP} \ \mathrm{(k \Omega)}$ & Resistance in AP state &  $R_\mathrm{P}\mathrm{(1+ TMR)}$\\
                                      &                        &  $= 22.5$\\
\hline
$\eta$ & Spin polarization & $\mathrm{\frac{\sqrt{TMR(2+TMR)}}{2(1+TMR)}}$ \\
       & efficiency        & $= 0.363$ \\
\hline
\end{tabular}
\label{tab:param}
\end{center}
\end{table}

\begin{figure}[b!]
\centering
\includegraphics[width = \columnwidth]{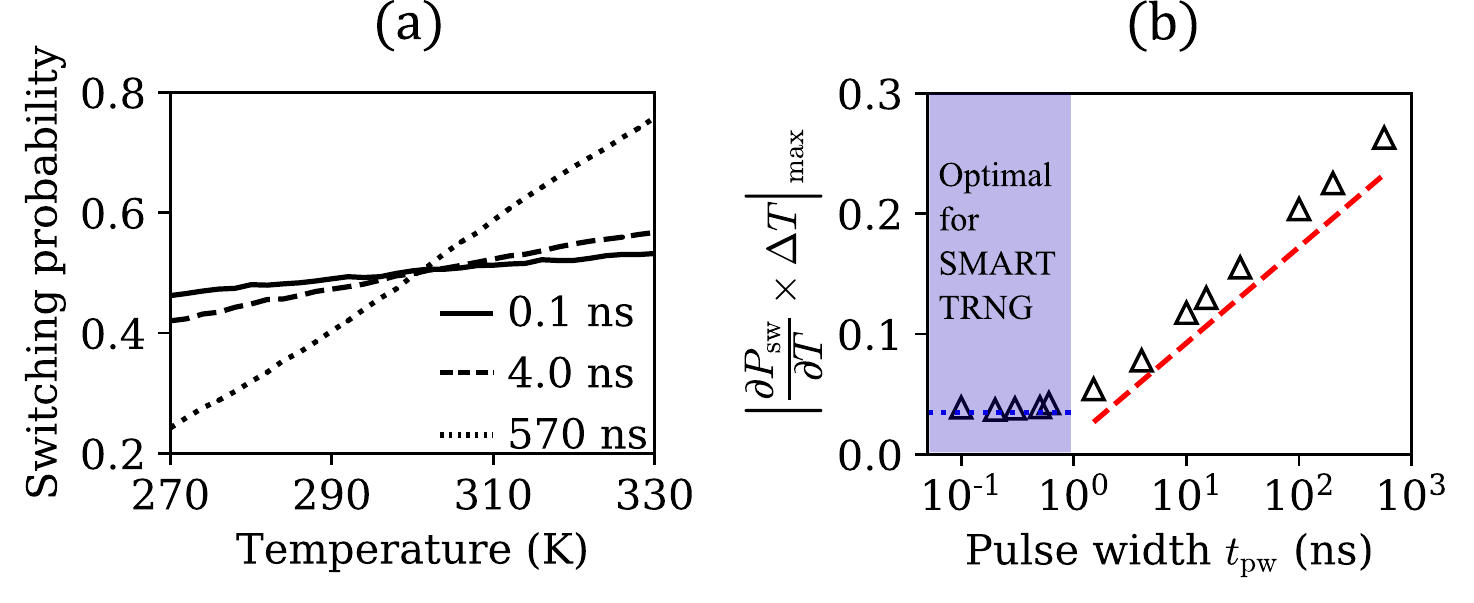}
\caption{(a) Switching probability as a function of temperature for different pulse widths. Increase in temperature leads to drift in the switching probability from the desired value of 0.5. This effect becomes prominent for higher pulse widths
(b) Maximum change in switching probability from the mean value of 0.5, for $T = 300 \pm 30$~$\mathrm{K}$, vs. pulse width. SMART TRNG will be operated in the short-pulse width regime (i.e., $t_\mathrm{pw}\leq 1$~$\mathrm{ns}$, also shown by the shaded region) where the switching probability is less dependent on temperature.} 
\label{fig:temp}
\end{figure}

\subsection{Effect of temperature on switching probability}
The current flowing through the tunnel junction can lead to an increase in its temperature due to Joule heating.
In addition, the ambient temperature of the chip that the TRNG is fabricated on could be different from that of the MTJ due to differential heating.
This would lead to design and reliability issues for the TRNG because the initial Boltzmann distribution changes with temperature. 
As a result, $P_\mathrm{sw}$ for a given applied voltage pulse would be different from the expected value of 0.5, as shown in Fig.~\ref{fig:temp}(a).
It can be observed that in all the cases, $P_\mathrm{sw}$ increases above 0.5 for temperatures above 300~$\mathrm{K}$ while it decreases below 0.5 for temperatures below 300~$\mathrm{K}$. Crucially, the change in $P_\mathrm{sw}$ with temperature is more prominent as $t_\mathrm{pw}$ increases. 
This suggests that the change in switching probability with temperature is higher for longer pulse widths (and lower pulse amplitude), the thermally-assisted switching regime~\cite{pinna2013}. This regime is, therefore, avoided in SMART TRNGs by operating the devices in the short-pulse ballistic limit.
As shown in Fig.~\ref{fig:temp}(b), the maximum switching probability change, $\left|\pdv{P_\mathrm{sw}}{T} \times \Delta T \right|_\mathrm{max}$, for a temperature variation of $\Delta T = \pm 30$~$\mathrm{K}$ around room temperature, is small in the short-pulse limit while it increases as the log of the pulse width in the long-pulse regime~\cite{rehm2023}. 
More specifically, 
$\qty|\pdv{P_\mathrm{sw}}{T}| \sim \frac{\log 2}{2 T}$ in the ballistic limit, as shown by the dotted blue line in Fig.~\ref{fig:temp}(b).
On the other hand, in the long-pulse diffusive limit $\qty|\pdv{P_\mathrm{sw}}{T}| \sim\frac{\log 2}{2 T} \qty(\log(\frac{\log 2}{\Gamma_0 t_\mathrm{pw}}))$, as shown by the red dashed line in Fig.~\ref{fig:temp}(b).
Here, $\Gamma_0 = 1$~GHz is the attempt frequency for nanomagnets~\cite{fukushima2014}. 
The strong temperature sensitivity of $P_\mathrm{sw}$ in the
diffusive limit stems from 
its double exponential dependence on energy barrier and temperature~\cite{fukushima2014, rehm2023}.

Overall, our Monte Carlo simulations confirm that, the operation of the SMART TRNG would be optimal for $V_\mathrm{ap} \in [0.41~\mathrm{V}, 0.54~\mathrm{V}]$ and the corresponding $t_\mathrm{pw} \in [0.48~\mathrm{ns}, 0.88~\mathrm{ns}]$. Herein, the variation in switching probability is small with both the pulse amplitude and duration, the energy of operation and the EDP are both low, and the effect of temperature variation is small.

\section{Statistical and NIST Test For Randomness}
To evaluate the statistical quality and randomness of the bitstream generated by the SMART TRNG at $T = 300~\mathrm{K}$, firstly, we use our stochastic LLG code to generate 10 different bitstreams of $2 \times 10^6$ bits each, corresponding to switching the FL from P to AP state with $P_\mathrm{sw} \approx 0.502$.

\begin{figure}[b!]
\centering
\includegraphics[width = \columnwidth]{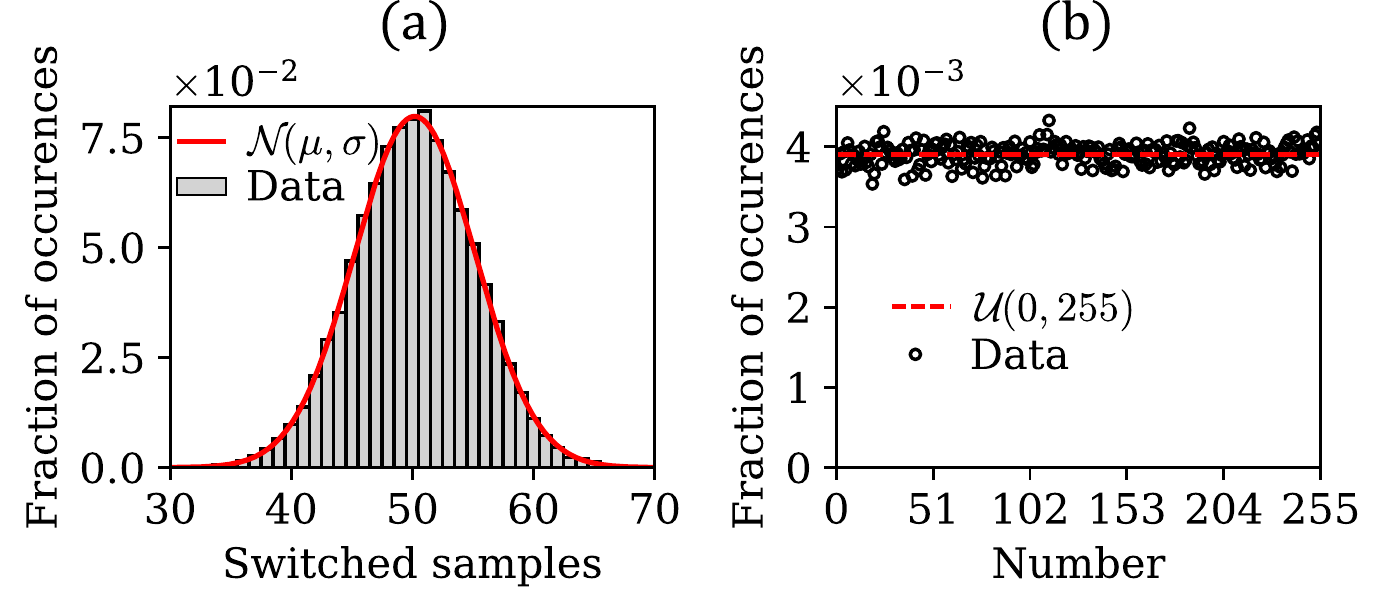}
\caption{(a) Distribution of the switched events from a sample of 100. The histograms were constructed by repeating this $2 \times 10^4$ times. The solid red line represents the normal distribution, $\mathcal{N} (50.2, 5)$.
(b) Distribution representing a number between 0 and 255 from a sample of $25 \times 10^4$. The dashed red line denotes uniform distribution, $\mathcal{U} (0, 255)$}
\label{statistical}
\end{figure}

Secondly, we choose one of the 10 bitstreams and create $2 \times 10^4$ non-overlapping random samples with $N = 100$ bits, each.
It is expected that the number of switched states, out of 100, in each sample would follow a binomial distribution with mean $\mu = N P_\mathrm{sw} = 50.2$ and standard deviation $\sigma = \sqrt{N P_\mathrm{sw} (1 - P_\mathrm{sw})} = 5$. 
This is shown in Fig.~\ref{statistical}(a) by the histograms while the red solid curve represents a normal distribution, $\mathcal{N} (\mu, \sigma)$.
Thirdly, we consider a set of eight contiguous bits of the $2 \times 10^6$ bits, and map them to a number between 0 and 255. 
Since the switching probability is almost 0.5, the probability of a `0' or `1' in the bitstream is almost the same, and, therefore, the probability of any number between 0 and 255 should follow uniform distribution, $\mathcal{U} (0, 255)$. 
Figure~\ref{statistical}(b) indicates that the numerically generated data (scattered symbols) roughly follow a uniform distribution (dashed red line).

Finally, we pass each bitstream through the National Institute of Standards and
Technology (NIST) test suite~\cite{bassham2010}, which consists of several frequency and non-frequency related tests.
We find that 2 out of the 10 raw bitstreams pass all 188 tests, whereas the rest 8 bitstreams pass 184 or 185 out of the 188 tests.  
The bitstream under consideration is said to have passed a certain test if the p-value of the testing operation is greater than 0.01~\cite{bassham2010}.
Failing a certain test implies that the bitstream is not truly random from the perspective of that test.
The common failed tests included the (monobit) frequency test, the forward and the reverse cumulative sum tests, and the run test.
On whitening the bitstream with just one XOR operation, 42 out of the 45 post-processed bitstreams pass all the 188 NIST tests while the other 3 pass 187 tests.
This is because an XOR operation enhances the entropy of the system by reducing the bias and tuning the switching probability closer to the ideal value of 0.5~\cite{fukushima2014, rangarajan2017}.
It must be noted that we used PRNGs to generate random white Gaussian noise when solving~(\ref{sLLGS}), which could have led to some correlations in the raw bitstreams generated by the SMART TRNG. 

\begin{figure}[b!]
\centering
\includegraphics[width = 0.8\columnwidth]{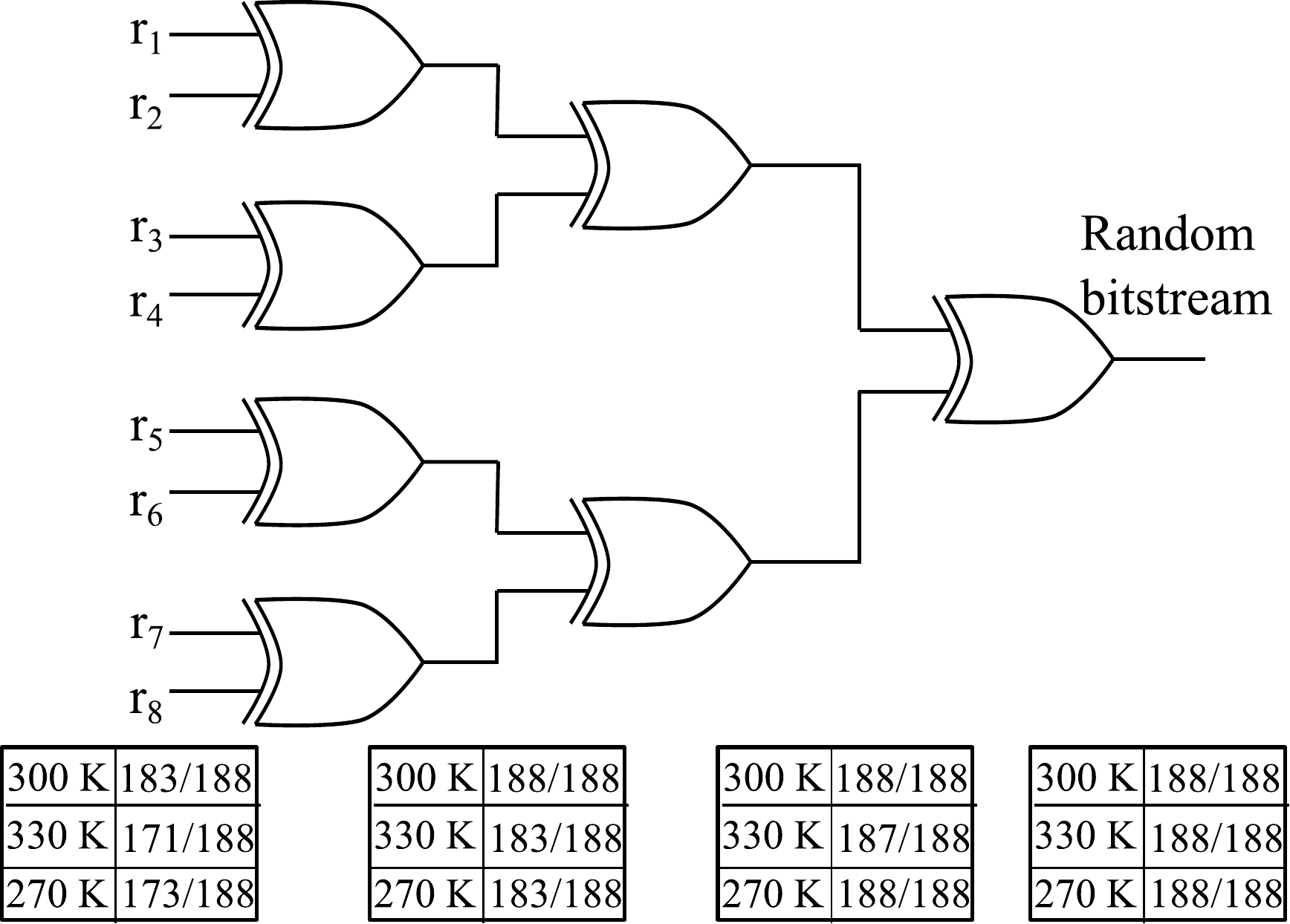}
\caption{XOR tree structure of TRNGs. Each r$_i$ is an individual TRNG. The number of passed NIST tests at each level for three different temperatures is listed. Here, we have reported the worst case scenarios.}
\label{fig:xor}
\end{figure}
An ideal switching probability of 0.5 can only be achieved if the input voltage pulse has exact amplitude and duration.
However, due to PVT variations, the switching probability could be different, therefore, post-processing the raw bitstream with at least one XOR operation might be a necessity.
Figure~\ref{fig:temp}(b) suggests that changes in the switching probability in the ballistic region is small, but present nonetheless.
Therefore, we check the randomness of the raw and XORed bitstreams generated by SMART TRNGs at different temperatures ranging from 270~$\mathrm{K}$ to 330~$\mathrm{K}$. Figure~\ref{fig:xor} summarizes the results of our analysis.
We find that the raw bitstreams fail several NIST tests at temperatures other than $300~\mathrm{K}$ since the switching probability has changed from its ideal value. However, when they are XORed their entropy is enhanced such that the resulting random output bitstream fails fewer NIST tests.
Further XORing the bitstream increases the randomness and the processed bitstream pass all NIST tests.

\begin{figure*}[ht!]
\centering
\includegraphics[width = 2\columnwidth]{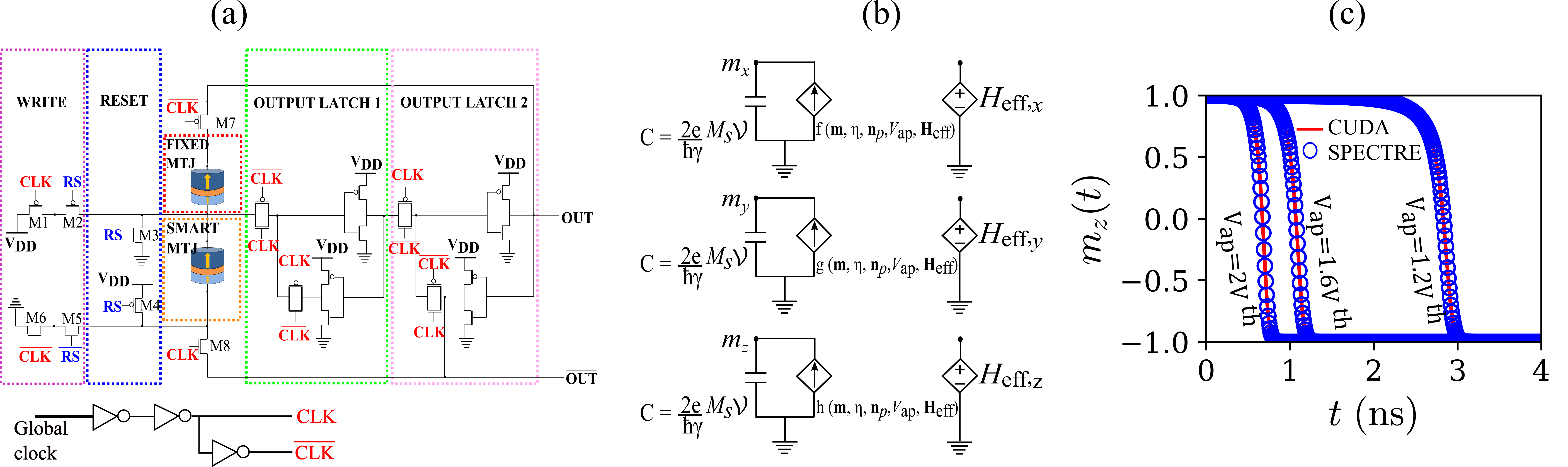}
\caption{(a) A circuit diagram for the TRNG described in this work. 
It consists of three parts---write, reset, and read. 
The write sub-circuit switches the MTJ with a probability of 50\% while the reset sub-circuit ensures that the MTJ is in parallel state before every write operation.
The read sub-circuit consists of two output latches and a reference MTJ.
The output state of the MTJ is digitized and collected at OUT terminal.
The circuit operation is controlled by two signals---RS and CLK.
(b) Equivalent circuit model for the FL of the SMART device. Detailed model in Ref.~\cite{bonhomme2014}.
(c) Comparison of z-component of $\vb{m}$ between CUDA and circuit simulator Cadence Spectre\textsuperscript{\textregistered} for different input voltages. For the purpose of comparison only, thermal noise was not included here.}
\label{fig:ckt}
\end{figure*}
\section{TRNG Circuit Design}\label{results}
A full SMART TRNG circuit, designed using 45-$\mathrm{nm}$ CMOS PDK~\cite{stine2007}, is presented in Fig.~\ref{fig:ckt}(a). 
It is composed of three main parts: write, reset, and read sub-circuits that are controlled by two external signals, viz. RS and CLK~\cite{mehri2022}. 
Here, BSIM4 Level 54 models were used for the transistors while a circuit-compatible SPICE model of a FM~\cite{bonhomme2014}, subject to thermal effects and spin-torque, was used for the FL of the SMART device.
As shown in Fig.~\ref{fig:ckt}(b), the magnetization of the FL, $\vb{m}$, in any direction, $j (= x, y, z)$, is modeled as the node voltage of a capacitor, whereas the effective field acting on the FM is represented as a voltage-controlled current source.
The results of this model were benchmarked against that of the LLG code implemented on CUDA. 
Though all the components of $\vb{m}$ match very well for different applied biases, we present in Fig.~\ref{fig:ckt}(c) only the z-component for the sake of brevity.
Finally, this model was integrated with the CMOS circuitry to obtain the full TRNG circuit, and the trapezoidal solver of Cadence Spectre\textsuperscript{\textregistered} with a time step of 10 fs was used for all transient simulations.

\begin{figure}[hb!]
\centering
\includegraphics[width = \columnwidth]{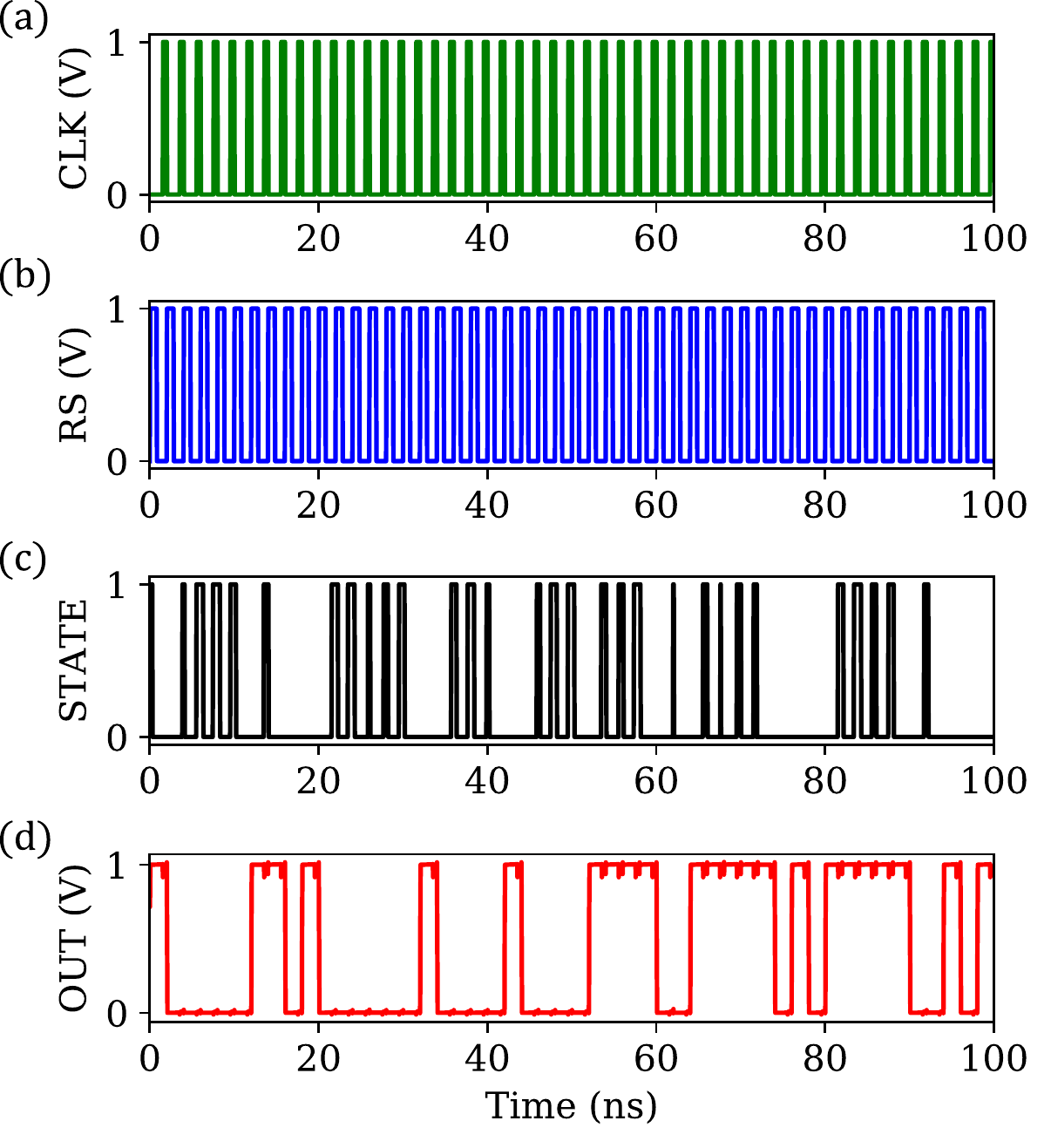}
\caption{ (a) CLK and (b) RS signals. Both of them have a time period of 2~$\mathrm{ns}$. RS is high for 0.8~$\mathrm{ns}$ whereas CLK is high for 0.46~$\mathrm{ns}$.
(c) Digitized state of the SMART MTJ.
(d) The voltage at the OUT node at the beginning of the next clock cycle depends on OUT during the current cycle and the state of the MTJ.
Reset phase: RS = 1, CLK = 0. The SMART MTJ is deterministically set to the P state, or STATE = 0.
Write phase: RS = 0, CLK = 0. Probabilistic switching of the SMART device from P state to AP state, or to STATE = 1.
Read phase: RS = 0, CLK = 1. Output of first latch is updated based on current value of OUT and STATE. Output of the second latch is updated in the next clock cycle.}
\label{fig:operation}
\end{figure}

Our SMART TRNGs are triggered by a short voltage-pulse, which leads to a stochastic switching of the MBM FL from parallel to antiparallel state. As a result, the SMART device should be initialized such that the FL and PL are in the P state before every write cycle. 
This is ensured by the reset circuit of Fig.~\ref{fig:ckt}, which comprises transistors M3 and M4 along with the SMART device, and is activated for RS = 1, CLK = 0.
In this work, the time period of both these signals, shown in Fig.~\ref{fig:operation}(a, b), is chosen as 2~$\mathrm{ns}$.
Here, RS is in state `1' for 0.8~$\mathrm{ns}$ while CLK is in state `1' for 0.46~$\mathrm{ns}$.
Next, in our design we fix the widths of both M3 and M4 to 180~$\mathrm{nm}$.
During this phase, current flows from FL to PL and a voltage drop of 0.8~$\mathrm{V}$ is developed across the SMART MTJ. This voltage is large enough to deterministically switch the FL to its P state.

The write sub-circuit of Fig.~\ref{fig:ckt}, which consists of transistors M1, M2, M5, and M6 along with the SMART MTJ, is activated when RS = 0, CLK = 0.
It provides the triggering voltage, with opposite polarity to that applied during the reset operation, to probabilistically switch the FL from its P state to AP state. 
In our design, we fix the width of M1 and M2 to 140~$\mathrm{nm}$ while that of M5 and M6 to 90~$\mathrm{nm}$. This leads to a voltage drop of 0.6~$\mathrm{V}$ across the SMART MTJ which ensures stochastic switching behavior. The digitized state of the MTJ is shown in Fig.~\ref{fig:operation}(c). State `1' (`0') corresponds to AP (P) relative orientation of the FL and PL.  

Finally, the read sub-circuit is activated for RS = 0, CLK = 1.
It extracts the random uncorrelated bitstream as a voltage signal measured at the output node, OUT, as shown in Fig.~\ref{fig:operation}(d). It comprises a fixed MTJ with fixed resistance ($R_\mathrm{P}< R < R_\mathrm{AP}$), transistors M7, M8, and two output latches. The latches serve as internal post-processing units and increase the entropy of the system. This is because the output of the first latch corresponds to an XOR operation between the state of the SMART MTJ (`0' or `1') and the output of the second latch (OUT)~\cite{mehri2022}. For example, if the SMART MTJ is in `0' (`1') state and OUT = `1', the voltage between the reference MTJ and the SMART MTJ would be lower (higher) than V$_\mathrm{DD}/2$. This would ensure that the output of the first latch is `1' (`0'). On the other hand, if the SMART MTJ is in `0' (`1') state and OUT = `0', the voltage between the reference MTJ and the SMART MTJ would be higher (lower) than V$_\mathrm{DD}/2$. This would lead to `0' (`1') at the output of first latch.
The output of the second latch is updated in the next clock cycle, as {OUT}$_{i+1}$ = $\overline{\mathrm{STATE} \oplus \mathrm{OUT}}_{i}$.
The can be easily observed from Fig.~\ref{fig:operation}(c, d).
Here, the resistance of the reference MTJ is assumed to be 18~$\mathrm{k \Omega}$. 
Table~\ref{tab:performance} lists the performance metrics of our SMART TRNG and compares it to prior MTJ-based TRNGs. We note that the EDP of the SMART TRNG is about 46\% of that of~\cite{rangarajan2017}.

\begin{table}[h!]
\caption{Comparison of performance metrics of the SMART TRNG to other MTJ-based TRNGs at T = 300~$\mathrm{K}$.}
\begin{center}
\begin{tabular}{|l|c|c|r|}
\hline
\textbf{TRNG} & \textbf{Bitrate}& \textbf{Energy} &  \textbf{Energy Delay}   \\
              & \textbf{$\mathrm{(Mb/s)}$} & \textbf{($\mathrm{fJ/bit}$)} & \textbf{Product ($\mathrm{J \cdot s}$)} \\
\hline
Spin dice~\cite{fukushima2014} & 0.6 &  122~\cite{rangarajan2017} & 0.2$\times 10^{-18}$\\
\hline
Precessional switching & 244 & 100 & 4$\times 10^{-22}$\\
 TRNG~\cite{rangarajan2017} & & &\\
\hline
Parallel TRNGs~\cite{qu2018} & 66.7-177.8 & 640-810 & 1-4.6$\times 10^{-21}$\\ 
\hline
LBM-TRNG~\cite{vodenicarevic2017} & $<$ 0.1 & 20 & $>$ 0.2$\times 10^{-18}$\\
\hline
SMART-TRNG (this work) & 500 & 92 & 1.84$\times 10^{-22}$\\
\hline
\end{tabular}
\label{tab:performance}
\end{center}
\end{table}

\section{Discussion}
The SMART TRNG circuit presented here uses a PMA FL (Fig.~\ref{fig1} (a)) with an energy barrier of $35 k_B T$, which is referred to as an MBM (i.e., medium-barrier magnet), operated in the short-pulse ballistic limit.
A direct consequence of operating in the ballistic regime is the reduced sensitivity of switching probability to temperature variations (Fig.~\ref{fig:temp}). 
This is because, in this regime of operation, the magnetization dynamics is strongly dependent on the torque due to input voltage pulse as compared to that due to thermal field.
On the contrary, in the long-pulse diffusive regime, the torque due to thermal field becomes stronger leading to a significantly higher dependence of switching probability on temperature variations (Fig.~\ref{fig:temp}).
Previous spintronic TRNGs were designed to operate in the diffusive limit~\cite{fukushima2014}, and therefore exhibited significant variation to temperature fluctuations~\cite{rehm2023}.
In addition, compared to diffusive operation, ballistic operation of the TRNG leads to lower energy consumption and lower energy-delay product, as shown in Fig.~\ref{energy_delay}.

The use of MBMs in the SMART MTJ reduces the overall energy required to switch the FL as the energy between the two stable states (P and AP) is lower than that in case of HBMs, which are typically used for NV memories.
For example, if the barrier height of the FL increases by $1.5\times$ due to an increase in its volume, the switching energy would roughly increase by $2.25\times$. 
This is because, firstly, the threshold voltage, $V^\mathrm{th} (\propto \mathcal{V})$, increases by $1.5\times$, which requires an increase in the input voltage amplitude, $V_\mathrm{ap}$.
Secondly, in the ballistic regime, the switching probability is dependent on the net charge~\cite{liu2014}, which is $G_p \qty(V_\mathrm{ap} - V^\mathrm{th}) t_\mathrm{pw}$.
This leads to the scaling of the switching time as $\sim \frac{(1 + \alpha^2)}{\alpha \gamma \mu_0 H_k \qty(V_\mathrm{ap}/V^\mathrm{th} - 1)} \log(4 \sqrt{\pi \frac{E_b}{k_B T}})$~\cite{sun2016, rehm2019}. 
Therefore, the switching energy, $G_p V_\mathrm{ap}^2 t_\mathrm{pw}$, scales as $\sim G_p V_\mathrm{ap}^2\frac{(1 + \alpha^2)}{\alpha \gamma \mu_0 H_k \qty(V_\mathrm{ap}/V^\mathrm{th} - 1)}\log(4 \sqrt{\pi \frac{E_b}{k_B T}})$.
Here, it is assumed that all the other material parameters including the resistance of the MTJ are exactly the same.
The above result suggests that FMs with lower energy barriers are preferable for reducing the switching energy. 
However, lowering the energy barrier by reducing the diameter of the FM could lead to fabrication-related reliability issues. 
Alternatively, the energy barrier could also be lowered by using ferromagnetic materials with lower effective anisotropy, and hence $H_k$. This would, however, increase the switching time since it scales inversely with $H_k$, as stated above.
Superparamagnets with $E_b \lesssim 5 k_B T$ have recently been shown to operate in the nanosecond regime~\cite{hayakawa2021, safranski2021}, and could be used as a fast and low energy source of random bitstreams~\cite{safranski2021}. Their operation, however, is mainly dominated by thermal noise, which leads to extreme sensitivity to temperature changes~\cite{rehm2023}. In addition, changes in $E_b$ due to any variation in material parameters or external magnetic perturbations could also affect their switching probability significantly~\cite{rehm2023}.

In this work, the bitstreams generated using the SMART MTJ device required at least one XOR operation in order to pass all the NIST tests even at $300~\mathrm{K}$ (Fig.~\ref{fig:xor}). 
This, in turn, requires at least two SMART MTJs and an XOR circuit, which increases the area and power overheads. 
The internal post-processing unit of the TRNG circuit considered in this work (Fig.~\ref{fig:ckt}(a)) enables this XOR operation internally, without the need for an external XOR circuit, at a lower energy and area cost~\cite{mehri2022}. 
As a result, the generated bitstreams pass all the 188 NIST tests at $300~\mathrm{K}$. However, if the temperature changes by $30~\mathrm{K}$ external XOR operations would be required in order to ensure that the bitstreams pass all NIST tests ( Fig.~\ref{fig:xor}).
The switching probability of the SMART MTJ is sensitive to the input voltage pulse width and amplitude (Figs.~\ref{pr_heatmap} and~\ref{pr}). The XOR operation alleviates this sensitivity. 

One of the drawbacks of the SMART TRNG is the necessary requirement of the reset operation. It increases the area and energy cost while decreasing the throughput.
Another key problem is the charge current flowing through the MTJ during both the reset and write operation.
The charge current adds stress on the tunnel barrier and makes it vulnerable to dielectric breakdown, which can compromise the reliability of the TRNG circuit. 
This problem of device endurance could be addressed by the use of a device set-up which utilizes the spin-orbit torque (SOT) to switch the magnetization of the FL. 
Towards this a bilayer of in-plane ferromagnet (CoFeB) and non-magnet (Ti) could be used to generate interfacial SOT with spin polarization perpendicular to the interface of both the FL and the bilayer~\cite{ryu2020}.
In this set-up, a small charge current would flow through the MTJ only during the read process, thereby increasing its endurance.

\section{Conclusion and Outlook} \label{conclusion}
In this work, we presented the design of a SMART (stochastic magnetic actuated random transducer) MTJ device utilizing an MBM magnet with PMA which serves as the device's free layer. Unlike LBM-based spintronic TRNGs, SMART TRNGs are faster with lower EDP and can be fabricated more easily.
Short voltage pulses in the ballistic limit were applied to the SMART device in order to tune its switching probability to 50\%. Through numerical simulations we obtained corresponding values of pulse amplitude and duration that would lead to 50\% switching probability. 
The effect of temperature on the switching probability was also found to be the lowest in the ballistic regime of operation. 
An optimal regime for low energy operation of SMART TRNGs was also suggested.
In general, bitstreams generated by the simulation of the SMART device required two levels of XOR operation to pass all NIST tests. 
A TRNG circuit with an internal post-processing XOR unit was considered. 
Our simulations showed that, as compared to previous spintronic TRNGs, the SMART TRNG circuit had a higher bitrate and relatively low energy and area costs. This is due to its operation in the ballistic mode and the use of the internal post-processing unit. 
For the material parameters considered in this work, uniformly distributed bitstreams comprising $0.5 \times 10^9$ bits can be generated at the cost of $46~\mathrm{\mu W}$ of power. 
Such bitstreams would be useful in Monte Carlo simulations, and as initial cipher-key in cryptography algorithms or for watermarking in other hardware primitives. 
By tuning the input pulse width or amplitude, bitstreams with different switching probabilities can also be generated for use in error-resilient and energy efficient stochastic computing tasks such as Bayesian inference.
Decoupling the read and write paths via the use of interfacial spin-orbit torque could improve the reliability of the TRNG circuit.

\section*{Acknowledgments}
The authors at UIUC acknowledge the support of NSF through Award \# CCF-1930620 and Air Force Research Laboratory under Grant \# FA8750-21-1-0002. 
The research at NYU was supported by the DOE Office of Science (ASCR/BES) Microelectronics Co-Design project COINFLIPS.
The research at UVA was supported by NSF I/UCRC on Multi-functional Integrated System Technology (MIST) Center; IIP-1439644, IIP-1439680, IIP-1738752, IIP-1939009, IIP-1939050, and IIP-1939012.

\end{document}